# Dithered quantizers with negligible in-band dither power

Abhishek Ghosh, *Student Member, IEEE* and Sudhakar Pamarti, *Member, IEEE*
Department of Electrical Engineering, University of California, Los Angeles, CA 90095
Email:{abhishek,spamarti}@ee.ucla.edu

*Abstract*—Subtractive dithered quantizers are examined to minimize the signal-band dither power. The design of finite impulse response(FIR) filters that shape most of the dither-power out of the signal band while maintaining the benefits of dithering are dealt with in detail. Simulation results for low-medium resolution quantizers are presented to highlight the overall design consideration.

## I. INTRODUCTION

Quantizers are the portals to digital signal processing of all real-world signals and hence serve as the main interface between natural and machine-based signal processing. The main purpose of a quantizer is to represent signals in a form that is easily operable, easy to store in digital computers. An example mid-tread quantizer is shown in Fig. 1. The quantizer is said to not *overload* if $|z[n]| \leq Q\Delta/2$ (note that throughout this paper we shall not make any distinction between signals $z[l]$ and $z_l$ where $l$ denotes the time-index) where $Q$ denotes the number of output levels (here $Q = 5$). As can be seen from Fig. 1, the input-output characteristic of any example quantizer, is evidently non-linear and hence signals when *quantized* produce spectral content hitherto absent in them [1]–[3], [5]. Of particular interest are sinusoidal signals [3], which are composed of discrete tonal frequencies. Such signals when passed through quantizers give rise to spurious tones corrupting the output signal spectrum. There is a rich body of literature attempting to find the resulting quantization error statistics and spectrum [1]–[3], [5]. A major understanding from all these works is that the input signal to the quantizer needs to be equipped with certain statistical properties in order to ensure that the quantization error samples are independent and uniformly distributed, a consequence of the latter being the ubiquitous $\Delta^2/12$ ($\Delta$ being the quantization step size) quantization error power. In most practical scenarios though, it is highly infeasible to handle signals with the required statistical properties (in fact for behavioral simulations, deterministic input signals are considered,viz. sinusoids which render the quantization error completely deterministic, given the quantizer characteristic). So, intuitively, a small signal, random in nature (called dither) is added to the input in order to make the composite signal samples *unpredictable* at a given time. In other words, the composite signal is constrained to possess the statistical properties outlined in [3]. Let us take a more formal view of quantization after dithering.

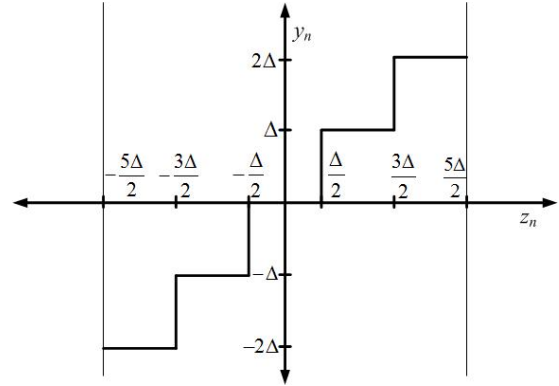

Fig. 1: Mid-tread quantizer

### A. Dithered quantization

A random signal $r[n]$ is added to the signal to be quantized $x[n]$ and the composite signal $z[n] = x[n] + r[n]$ is passed through the quantizer as shown in Fig. 2. There is a subtle difference though, between Fig. 2(a) and (b). In Fig. 2(b), the added dither signal $r[n]$ is subtracted digitally from the quantized value $y[n]$ and hence is called a *sub-tractively dithered quantizer*. Likewise, Fig. 2(a) refers to a non-subtractively dithered quantizer (commonly phrased as additive dithered quantizer). The added dither, $r[n]$ is usually constrained to be bounded between one least significant bit (LSB) of the quantizer. Separate conditions [3] have been theoretically derived for either case to ensure that the error-samples ($e[n] = y[n] - x[n]$ for Fig. 2(a) and $e[n] = y[n] - z[n]$ for Fig. 2(b)) are independent and uniformly distributed both in terms of first and second order statistics, formally

- $e_n$ is uniformly distributed.
- $(e_n, e_{n-p})$ are pairwise independent and uniformly distributed $\forall p \in \mathbb{Z} - (0)$.
- $e_n$ is independent of $x_{n-m} \forall m \in \mathbb{Z} - (0)$.

From the conditions outlined in [3], it becomes evident that the properties listed above are a lot more likely to hold for a subtractively dithered quantizer than an additive one. In fact, for the latter, it can be shown that the error samples $e[n]$ are never statistically independent of the input samples $x[n-p] \forall p \in \mathbb{Z}$. On the other hand, for a subtractively dithered quantizer, the main conditions imposed on the added dither

$r[n]$ for the above conditions to hold are:

$$\Phi_{r_n}(u)_{|u=\frac{k}{\Delta}} = 0$$
$$k \in \mathbb{Z} - (0)$$
$$\Phi_{r_n,r_{n-p}}(u_1,u_2)_{|u_1=\frac{k_1}{\Delta},u_2=\frac{k_2}{\Delta}} = 0$$
$$k_1, k_2 \in \mathbb{Z}^2 - (0,0) \quad (1)$$

where $\Phi_w(u)$ is the characteristic function(cf) of the random variable $w$ and $\Phi_{w_1,w_2}(u_1,u_2)$ is the joint characteristic function(jcf) of random variables $w_1$ and $w_2$.

Unfortunately, such $r[n]$ would contribute too much noise to the quantizer output. In fact, a uniformly distributed dither signal would degrade the overall signal-to-noise ratio(SNR) by 3dB. Furthermore, it may be impossible or at least extremely challenging to digitally generate such dither. The immediate solution to such a problem is to spectrally shape the dither energy out of the signal band of interest [4]. Such an architecture is presented in Fig. 2(c) as an extension of Fig. 2(b). However, filtering a signal tantamounts to modifying its statistical properties. Consequently, the filtered signal $r[n]$ in Fig. 2(c) may not possess the properties outlined in (1). There have been some very interesting works treating filtered dither signals and their efficacies in whitening the quantization error, notable among which are [4], [6], [7]. With reference to Fig. 2(c), in [4], a detailed analysis is done on the properties of $r[n]$ where $d[n]$'s are i.i.d. random variables. However, the analysis is specific to additive dithered quantizers and imposes very strict conditions on the filter-coefficients (FIR or IIR). In [7], a simplified condition is derived for FIR filters to ensure that the error-samples possess the properties outlined in (1). However, the quantizer treated in [7], works on integer values only and also the whitening conditions are derived only for error-sample pair that are apart from each by at least the filter length. The work in [6] also provides conditions for the impulse response of the IIR filter (integrator in feed-forward path of a sigma-delta modulator) to ensure (1). In this work, we provide an alternative technique to digitally filter (using a FIR filter) a bi-valued dither signal, which enables the dither signal to span a finite number of values within the coarse LSB pushing most of the dither energy to frequencies where the input signal has no or negligible content. We theoretically derive conditions for achieving complete whitening of the error sequence. In the next section, we put forth the proposed technique and investigate conditions for (1) to hold. In Section III, we simplify some of the conditions from Section II to ensure *almost* whitening of the error signal for a continuous-valued error-signal. In Section IV, we furnish pertinent simulation results to support our claim and we conclude the paper in Section V.

## II. Proposed Technique

With reference to Fig. 2(c), let us define a Bernoulli sequence $d[n]$ that follows the statistics:$\mathbb{P}r(d[n] = 0) = \mathbb{P}r(d[n] = 1) = 0.5$. The sequence $d[n]$ is passed through a digital filter $G(z)$ having a finite impulse response $g[n]$ of length $K$ to produce an output $r[n]$. The filter gain is so adjusted that the output $r[n]$ is in $[-\Delta/2, \Delta/2]$. Consequently, the filtered output $r[n]$ can be expressed as

$$r[n] = \frac{\Delta}{L}(g[0]d[n] + g[1]d[n-1] + \cdots + g[K-1]d[n-K+1]) \quad (2)$$

where $L$ is the $\mathcal{L}_1$ norm of the filter $g$. The quantity $\Delta/L$ can be thought of as the dither LSB (the minimum resolution of the added signal $r_n$).

*Note*: It should be pertinent to observe here that since the added filtered dither signal $r_n$ has a finite resolution, namely $\Delta/L$, hence any input signal $x_n$ below this resolution will not experience any whitening action. In the following arguments, we shall assume that the input signal $x_n$ is sufficiently large than the dither LSB so that any correlation arising due to the input signal residing between the dither steps, is negligible.

We propose the following theorem to ensure an *almost white* error sequence.

*Theorem 1:* Suppose the input to a non-overloading $Q$-level quantizer is $z_n = x_n + r_n$ where $r_n = g_n * d_n$ and $x_n$ is a bounded sequence i.e. $x_n \in [-(Q-1)\Delta/2, (Q-1)\Delta/2]$ for a sample mid-tread quantizer. Let $(U, V)$ be two independent, uniformly distributed random variables in $(-\Delta/2, \Delta/2)$. For all $(k_1, k_2) \in (-L/2, L/2]$

i) $e_n$ is independent and identically distributed uniformly

ii) $(e_n, e_{n-p})$ converges in distribution to $(U, V)$ $\forall p \in \mathbb{Z} - (0)$ iff at least one of the following conditions hold:

1) A non-negative integer $l < p$ exists such that $\langle g_l k_1 \rangle_L = L/2$

2) A non-negative integer $1 \leq r \leq p$ exists such that $\langle g_{K-r} k_2 \rangle_L = L/2$

3) A non-negative integer $p \leq m < K$ exists such that $\langle g_m k_1 + g_{m-p} k_2 \rangle_L = L/2$

where $\langle \rangle_T$ operator denotes modulo-$T$ operation.

*Remark*: We shall prove Property (ii) above to derive conditions (1), (2) and (3) and then lead to the proof of property (i) as a simplified subset.

*Proof:* The proof would use characteristic functions to derive conditions on the specific properties of the added dither signal. This is a commonly used technique for such applications [7]. In fact, from [3], we know, that the joint characteristic function for error-samples $(e_n, e_{n-p})$ can be written as

$$\begin{aligned}\Phi_{e_n,e_{n-p}}(u_1,u_2) = &\sum_{k_1=-\infty}^{\infty}\sum_{k_2=-\infty}^{\infty} \frac{\sin(\pi\Delta(u_1 - k_1/\Delta))}{(\pi\Delta(u_1 - k_1/\Delta))} \\ &\frac{\sin(\pi\Delta(u_2 - k_2/\Delta))}{(\pi\Delta(u_2 - k_2/\Delta))} \\ &\Phi_{x_n,x_{n-p}}(\frac{-2\pi k_1}{\Delta}, \frac{-2\pi k_2}{\Delta}) \\ &\Phi_{r_n,r_{n-p}}(\frac{-2\pi k_1}{\Delta}, \frac{-2\pi k_2}{\Delta}) \end{aligned} \quad (3)$$

Hence, for the joint density of $(e_n, e_{n-p})$ to converge to

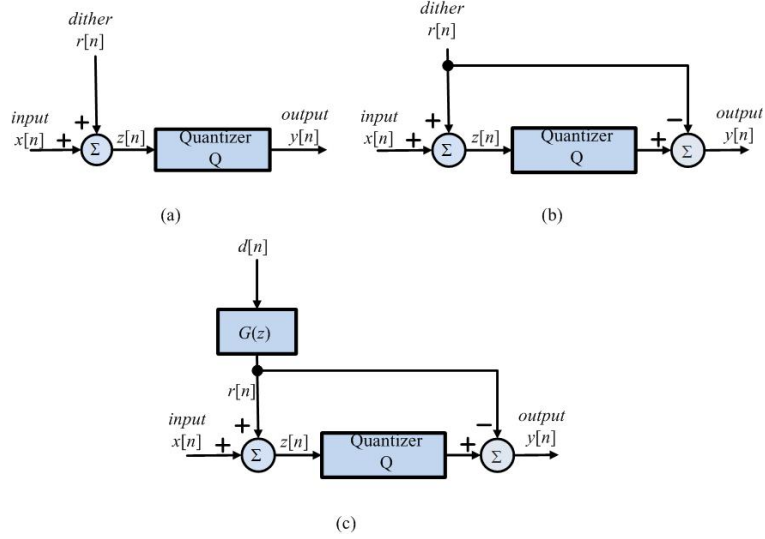

Fig. 2: Dithered Quantizers: (a) Non-subtractive/Additive (b) Subtractive (c) Filtered-subtractive

$(U, V)$, it suffices to show [3],

$$\Phi_{r_n, r_{n-p}}\left(\frac{-2\pi k_1}{\Delta}, \frac{-2\pi k_2}{\Delta}\right) = 0$$
$$\forall (k_1, k_2) \in \mathbb{Z}^2 - (0, 0) \quad (4)$$

Now, since $r[n]$ takes on only finite values in the set $\mathcal{A} = (-\Delta/2, -\Delta/2 + \Delta/L, -\Delta/2 + 2\Delta/L, \ldots, \Delta/2]$, we can write

$$\Phi_{r_n, r_{n-p}}(u_1, u_2) \triangleq \int_{r_1=-\infty}^{\infty} \int_{r_2=-\infty}^{\infty} f_{r_n, r_{n-p}}(r_1, r_2) e^{-u_1 r_1 - u_2 r_2} dr_1 dr_2$$

$$= \sum_{m_1=-L/2+1}^{L/2} \sum_{m_2=-L/2+1}^{L/2} \Pr\left(r_n = m_1 \frac{\Delta}{L}, r_{n-p} = m_2 \frac{\Delta}{L}\right) e^{-j(u_1 m_1 + u_2 m_2)\frac{\Delta}{L}} \quad (5)$$

Substituting in (5),

$$\Phi_{r_n, r_{n-p}}\left(\frac{-2\pi k_1}{\Delta}, \frac{-2\pi k_2}{\Delta}\right) = \sum_{m_1=-L/2+1}^{L/2} \sum_{m_2=-L/2+1}^{L/2} \Pr\left(r_n = m_1 \frac{\Delta}{L}, r_{n-p} = m_2 \frac{\Delta}{L}\right) e^{j\frac{2\pi(k_1 m_1 + k_2 m_2)}{L}}$$

$$\equiv \Phi_{r_n, r_{n-p}}\left(-\frac{2\pi k_1}{L}, \frac{-2\pi k_2}{L}\right) \quad (6)$$

Clearly, the RHS of (6) is $L$-periodic in $(k_1, k_2)$. In essence, $L^2$ number of jcf's should be accounted for, to ensure the condition given in (4) is true.

$$\Phi_{r_n, r_{n-p}}(u_1, u_2) = \mathbb{E}(e^{j(u_1 r_n + u_2 r_{n-p})})$$
$$= \mathbb{E}(e^{j(u_1 \sum_{m=0}^{K-1} g_m d_{n-m} + u_2 \sum_{l=0}^{K-1} g_l d_{n-p-l})})$$
$$= \prod_{l=0}^{p-1} \Phi_d(u_1 g_l) \prod_{m=p}^{K-1} \Phi_d(u_1 g_m + u_2 g_{m-p})$$
$$\prod_{r=1}^{p} \Phi_d(u_2 g_{K-r}) \quad (7)$$

For a Bernoulli dither $d_n$, with $\Pr(d_n = 0) = \Pr(d_n = 1) = 0.5$, $\Phi_d(v)$(cf of $d_n$)$= e^{(-jv/2)} \cos(v/2)$. Thus, we can write,

$$|\Phi_{r_n, r_{n-p}}(u_1, u_2)| = \prod_{l=0}^{p-1} |\cos(\frac{u_1 g_l}{2})| \prod_{m=p}^{K-1} |\cos(\frac{u_1 g_m + u_2 g_{m-p}}{2})|$$
$$\prod_{r=1}^{p} |\cos(\frac{u_2 g_{K-r}}{2})|$$

$$|\Phi_{r_n, r_{n-p}}\left(-\frac{2\pi k_1}{L}, \frac{-2\pi k_2}{L}\right)| = \prod_{l=0}^{p-1} |\cos(\frac{\pi k_1 g_l}{L})|$$
$$\prod_{m=p}^{K-1} |\cos(\frac{\pi(k_1 g_m + k_2 g_{m-p})}{L})|$$
$$\prod_{r=1}^{p} |\cos(\frac{\pi k_2 g_{K-r}}{L})| \quad (8)$$

So, based on conditions (1)-(3), the above result goes to zero $\forall (k_1, k_2) \neq (0, 0)$. ∎

The proof for condition (i) follows on similar lines as above. For the sake of completeness, we proceed as shown.

*Proof:* i) From [1],

$$\Phi_{e_n}(u) = \sum_{k=-\infty}^{\infty} \Phi_{z_n}(\frac{-2\pi k}{\Delta}) \frac{\sin(\pi\Delta(u-k/\Delta))}{\pi\Delta(u-k/\Delta)}$$
$$= \sum_{k=-\infty}^{\infty} \Phi_{x_n}(\frac{-2\pi k}{\Delta}) \Phi_{r_n}(\frac{-2\pi k}{\Delta}) \frac{\sin(\pi\Delta(u-k/\Delta))}{\pi\Delta(u-k/\Delta)} \quad (9)$$

Thus, for the LHS of (9) to converge to that of a uniform random variable, it is sufficient to show that

$$\Phi_{r_n}(\frac{-2\pi k}{\Delta}) = 0$$
$$\forall (k) \in \mathbb{Z} - (0) \quad (10)$$

From (6), it follows

$$|\Phi_{r_n}(\frac{-2\pi k}{\Delta})| = |\Phi_{r_n}(\frac{-2\pi k}{L})|$$
$$= |\sum_{m=-L/2+1}^{L/2} \Pr(g_0 d_n = m) *$$
$$\Pr(g_1 d_{n-1} = m) * \ldots * \Pr(g_{K-1} d_{n-K+1} = m)$$
$$\exp j \frac{2\pi k}{L} m|$$
$$= |\Phi_{d_n}(-\frac{2\pi k g_0}{L}) \Phi_{d_n}(-\frac{2\pi k g_1}{L}) \ldots$$
$$\Phi_{d_n}(-\frac{2\pi k g_{K-1}}{L})|$$
$$= \prod_{i=0}^{K-1} |\cos(\frac{\pi k g_i}{L})| \quad (11)$$

where * denotes the convolution operation.

From condition (1) (or (3)), letting $|p| \geq K$, one of the above product term goes to zero $\forall k \in \mathbb{Z} - (0)$. ∎

*Remarks*: From conditions (1)-(3) in Theorem 1, it may not be always possible to find FIR filter coefficients which imparts an appreciable in-band dither energy suppression as well as to satisfy the enumerated conditions. In the following section, we propose another theorem which ensures *almost* whiteness

## III. SIMPLIFIED CONDITIONS FOR APPROXIMATE WHITENESS

*Theorem 2:* Suppose the input to a $Q$-level non-overloading quantizer is $z_n = x_n + r_n$ where $r_n = g_n * d_n$ and $x_n$[1] is a bounded sequence i.e. $x_n \in [-(Q-1)\Delta/2, (Q-1)\Delta/2]$. Let $(U, V)$ be two independent, uniformly distributed random variables in $(-\Delta/2, \Delta/2]$.

i) $e_n$ is identical and independently (uniform) distributed and

ii) $(e_n, e_{n-p})$ converges in distribution to $(U, V)$ for all $|p| \geq K$ iff

1) The FIR filter coefficients $g[k]$ are of the form $2^i$ where $i \in [0, s-1]$ at least once and

---
[1]As noted before, it is assumed that $x_n$ is sufficiently greater than the dither LSB $\Delta/L$ in amplitude.

2) $L = ||g||_1 = 2^s$
where $s \in \mathbb{Z} > 1$

*Proof:*

Here also, we shall start from property (ii) and then lead to the proof for property (i). As pointed out in the proof for Theorem 1, in order to prove (ii), it suffices to prove

$$\Phi_{r_n,r_{n-p}}(\frac{-2\pi k_1}{\Delta}, \frac{-2\pi k_2}{\Delta}) = 0$$
$$\forall (k_1, k_2) \in \mathbb{Z}^2 - (0,0)$$

Again, from (6), this amounts to proving,

$$\Phi_{r_n,r_{n-p}}(\frac{-2\pi k_1}{L}, \frac{-2\pi k_2}{L}) = 0$$
$$\forall (k_1, k_2) \in \mathbb{Z}^2 - (0,0)$$

Since, $|p| \geq K$, hence it is not difficult to see that,

$$f_{r_n,r_{n-p}}(r_1, r_2) = f_{r_n}(r_1) f_{r_{n-p}}(r_2)$$
$$\Phi_{r_n,r_{n-p}}(u_1, u_2) = \Phi_{r_n}(u_1) \Phi_{r_{n-p}}(u_2) \quad (12)$$

Based on (11,12), one can see,

$$|\Phi_{r_n,r_{n-p}}(\frac{-2\pi k_1}{\Delta}, \frac{-2\pi k_2}{\Delta})| = |\prod_{i=0}^{K-1} \Phi_d(\frac{-2\pi k_1 g_i}{\Delta}) \Phi_d(\frac{-2\pi k_2 g_{K-1-i}}{\Delta})|$$
$$= \prod_{i=0}^{K-1} |\cos(\frac{\pi k_1 g_i}{L})||\cos(\frac{\pi k_2 g_{K-1-i}}{L})| \quad (13)$$

Now it becomes useful to consider the following cases, $\forall (k_1, k_2) \in [-L/2+1, L/2]$ assuming the conditions in Theorem 2 hold.

- $k_1 = odd, k_2 = odd$ One product term of the right-hand side of (13) can be written as $\cos(\pi k_j \frac{2^r}{2^s}), j = 1, 2$. Hence for $r = s-1$, we can write the product term as $\cos(\frac{\pi}{2} k_j)$ which goes to 0 since $k_{1,2}$ are odd.
- $k_1 = odd, k_2 = even$ Here, $k_1$ will drive the product term to 0 for $r = s-1$. The symmetric case of $k_2 = odd, k_1 = even$ similarly can be shown to equate to 0.
- $k_1 = even, k_2 = even$ Here, let $k_{1,2} = 2^l(2m+1), l \leq s-1$ for any integer $m$. Then the product term containing $r = s-1-l$ would yield $\cos(\frac{\pi}{2}(2m+1))$ which again goes to 0.

∎

Property (i) of Theorem 2 follows, almost directly, from the proof above, and hence is not given here for brevity.

## IV. SIMULATION RESULTS

We present here simulation results pertaining to the simplified conditions from Theorem 2, since as noted previously, it may not always be possible to impart appropriate high-pass shape to the dither signal satisfying the conditions of Theorem 1. Let us consider two example filters,

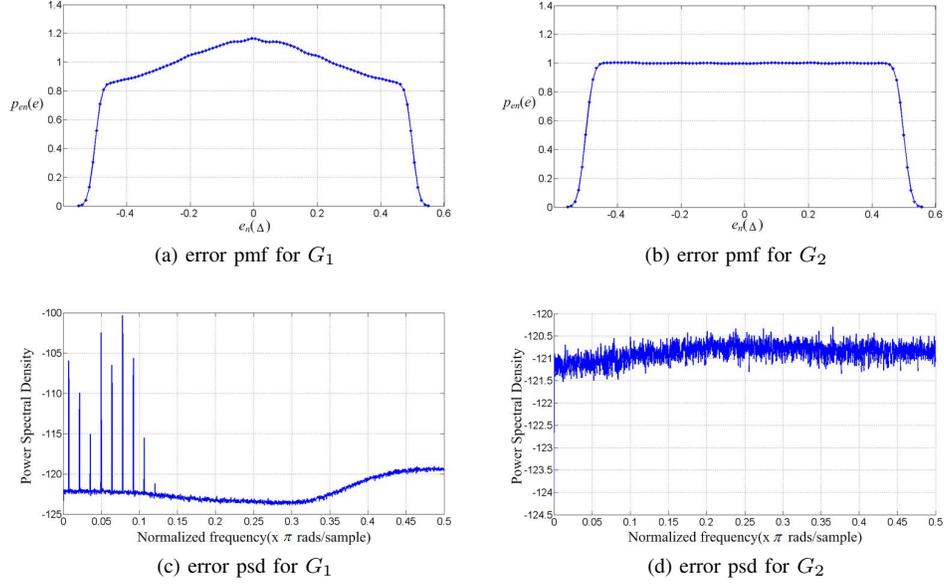

Fig. 3: Comparison of $G_1$ and $G_2$

$$G_1(z) = 1 - 3z^{-1} + 5z^{-2} - 9z^{-3} + 3z^{-4} - 3z^{-5} + 9z^{-6} - 5z^{-7} + 3z^{-8} - z^{-9}$$

$$G_2(z) = -1 - 2z^{-1} - 4z^{-2} - 8z^{-3} + 16z^{-4} - z^{-5}$$

Verifying, we find $G_1(z)$ satisfies neither of the two conditions of Theorem 2, while $G_2(z)$ satisfies both. The input $x[n]$ is chosen to be a sinusoid with an amplitude of $2\Delta$. The signal is quantized into $Q = 5$ levels as in Fig. 1. In Fig. 3(a),(b), we plot the pmf of the error sequence $e_n$ for both the cases, while Fig. 3(c)(d) shows the spectra of the error signal. As can be clearly seen, the proposed filter, namely $G_2$, whitens the error-sequence and exhibits an almost uniform pdf (Fig. 3(b)) while $G_1$ shows an almost triangular pmf (Fig. 3(a)) for the error samples. The power spectral densities also provide information to that end. The error psd for $G_2$ (Fig. 3(d)) is *white*, while the error psd for $G_1$ exhibits multiple spurious tones at harmonic frequencies (as is expected from a lookup table type non-linearity) (Fig. 3(c)). In order to make a fair comparison, a third case where a uniform dither signal $r[n]$ (the case in Fig. 2(b)) is added to the input signal before quantizing, is also considered. The spectra of $y[n]$ is plotted for all the three cases: $G_1, G_2$ and uniform dither in Fig. 4. As can be seen, the uniform dithered quantizer contributes the maximal in-band power while whitening the output spectrum completely. $G_2$ shapes the in-band dither power, as well as gets rid of any spurious components, while $G_1$ has the least in-band dither power contribution but engenders harmful spurious tones at the quantizer output.

## V. Conclusion

A dithering technique in quantizers is proposed. The technique relies on FIR filtering of the dither signal minimizing in-band SNR corruption. Theoretical conditions on the filter structure are derived to ensure whitening of the quantization error signal. Behavioral simulation results are presented to corroborate the proposed technique and claims.

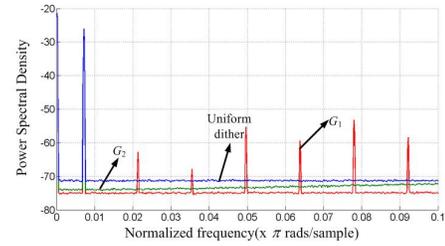

Fig. 4: Spectrum of $y[n]$ for three different scenarios